\documentclass{IEEEtran}

\usepackage[usenames,dvipsnames]{xcolor}
\usepackage{amsmath}
\usepackage{amssymb}
\usepackage{amsfonts}
\usepackage{cite}
\usepackage{epsfig}
\usepackage{array}
\usepackage{subfig}
\usepackage{pgf}
\usepackage{color}
\usepackage{algorithm}
\usepackage{float}
\usepackage{enumerate}
\usepackage{wrapfig}

\newtheorem{thm}{Theorem}

\newtheorem{lem}[thm]{Lemma}

\newcommand{\beq}{\begin{equation}}
\newcommand{\eeq}{\end{equation}}
\newcommand{\bea}{\begin{eqnarray}}
\newcommand{\eea}{\end{eqnarray}}
\newcommand{\bean}{\begin{eqnarray*}}
\newcommand{\eean}{\end{eqnarray*}}
\newcommand{\bit}{\begin{itemize}}
\newcommand{\eit}{\end{itemize}}
\newcommand{\ben}{\begin{enumerate}}
\newcommand{\een}{\end{enumerate}}
\newcommand{\blem}{\begin{lem}}
\newcommand{\elem}{\end{lem}}
\newcommand{\bthm}{\begin{thm}}
\newcommand{\ethm}{\end{thm}}
\newcommand{\bpf}{\begin{proof}}
\newcommand{\epf}{\end{proof}}

\newcommand{\sys}{0}
\newcommand{\supth}{\textrm{th}}

\begin{document}
\title{On Minimizing Data-Read and Download for Storage-Node Recovery}
\author{Nihar B. Shah\thanks{The author is with the department of Electrical Engineering and Computer Sciences, University of California, Berkeley. E-mail: nihar@eecs.berkeley.edu}}
\maketitle
\thispagestyle{empty}

\begin{abstract}
We consider the problem of efficient recovery of the data stored in any individual node of a distributed storage system, from the rest of the nodes. Applications include handling failures and degraded reads. We measure efficiency in terms of the amount of data-read and the download required. To minimize the download, we focus on the minimum bandwidth setting of the `regenerating codes' model for distributed storage. Under this model, the system has a total of $n$ nodes, and the data stored in \textit{any} node must be (efficiently) recoverable from \textit{any} $d$ of the other $(n-1)$ nodes. Lower bounds on the two metrics under this model were derived previously; it has also been shown that these bounds are achievable for the amount of data-read and download when $d=n-1$, and for the amount of download alone when $d\neq n-1$. 

In this paper, we complete this picture by proving the converse result, that when $d\neq n-1$, these lower bounds are strictly loose with respect to the amount of read required. The proof is information-theoretic, and hence applies to non-linear codes as well. We also show that under two (practical) relaxations of the problem setting, these lower bounds can be met for both read and download simultaneously.
\end{abstract}

\section{Introduction}
Consider a distributed storage system with $n$ storage nodes, each of which has a storage capacity of $\alpha$ bits. Data of size $B$ bits is to be stored across these nodes in a manner that the entire data can be recovered from any $k$ of the $n$ nodes. A problem that has received considerable attention in the recent past is that of efficient recovery of the data stored in an individual node, from the data stored in the remaining nodes in the system. This arises during handling of failures in distributed storage systems: upon failure of a node, it is replaced by a new node that must (efficiently) recover the data stored previously in the failed node from the remaining nodes in the system. A second application is that of degraded reads: if a node is busy or temporarily unavailable, then any request for the data stored in that node must be served (quickly) by downloading data from the remaining nodes.

We measure the efficiency of this process in terms of two metrics: the amount of data that must be read at the other nodes, and the amount of data downloaded from them. To optimize the amount of download, we consider the minimum bandwidth (MBR) setting of the regenerating codes model~\cite{YunDimKanJournal_short} for distributed storage. Under this model, recovery of the data stored in \textit{any} individual node must be accomplished by connecting to \textit{any} $d~(k\leq d<n)$ other nodes and downloading $\beta_D$ bits of data from each of them. Furthermore, under this model, these parameters must satisfy the condition
\beq d\beta_D=\alpha~. \label{eq:mbr_alpha_1}\eeq
An intuitive explanation of~\eqref{eq:mbr_alpha_1} is that recovery of the data stored in a node should entail only as much download as the amount stored. We shall assume throughout this paper that~\eqref{eq:mbr_alpha_1} is satisfied.

Under the MBR setting described above, a lower bound on the amount of download was derived in~\cite{YunDimKanJournal_short} as $\beta_D \geq \frac{B}{kd-{k \choose 2}}$. It is easy to see that the amount of data that is read at a node is at least as much as the amount of data downloaded from that node.~\footnote{The download may be smaller than the amount of read, since the data passed may be a (non-injective) function of the data that is read.} It follows that the total amount of read $\beta_R$ at any of the $d$ nodes helping in the recovery must obey $\beta_R \geq \beta_D$, and hence is also lower bounded as
$\beta_R \geq \frac{B}{kd-{k \choose 2}}$.

In this paper, we investigate the existence of codes that satisfy the aforementioned lower bounds with equality, i.e., satisfy
\bea \beta_D &=& \frac{B}{kd-{k \choose 2}}~\label{eq:mbr_eq_d}, \\
\beta_R &=& \frac{B}{kd-{k \choose 2}}~\label{eq:mbr_eq_r}\eea
for the recovery of the data of any of the $n$ nodes from any $d$ other nodes in the system. It was shown previously in~\cite{ourAllerton_supershort,ourAllertonJournal_short} that when $d=n-1$, the amount of download and read can simultaneously achieve~\eqref{eq:mbr_eq_d} and~\eqref{eq:mbr_eq_r} respectively for the recovery of the data of any individual node. Also, explicit codes with a download equalling~\eqref{eq:mbr_eq_d} for all values of the parameters were constructed previously in~\cite{ourProductMatrix_short}. However, it remained unknown whether or not the lower bound on the read~\eqref{eq:mbr_eq_r} can also be matched along with that on the download~\eqref{eq:mbr_eq_d} when $d \neq n-1$.

We complete this picture by showing that under the MBR setting described above, when $d\neq n-1$, it is \textit{impossible} to construct codes that simultaneously satisfy~\eqref{eq:mbr_eq_d} and~\eqref{eq:mbr_eq_r} for the download and read respectively. The proof is information-theoretic, and allows us to conclude that these bounds cannot be met even with non-linear codes. 

We also consider two (practical) relaxations of the problem setting, under which we provide explicit codes that can simultaneously achieve both~\eqref{eq:mbr_eq_d} and~\eqref{eq:mbr_eq_r} for all values of the system parameters. Under the setting described above, the data of \textit{any} individual node must be recoverable from \textit{any} $d$ other nodes, with the download and read satisfying~\eqref{eq:mbr_eq_d} and~\eqref{eq:mbr_eq_r} respectively. The two relaxations respectively weaken the two ``any'' criteria with respect to the read. Under the first relaxation, we require the read to achieve~\eqref{eq:mbr_eq_r} for only the recovery of the data stored in the systematic nodes (recovery of the data of the remaining nodes are allowed to have a larger read). This relaxed setting is relevant to the problem of degraded reads, where typically, the data stored in (only) the systematic nodes is of interest. Under the second relaxation, for the recovery of the data of any node, we require that~\eqref{eq:mbr_eq_r} be achieved for the read from at least one set of $d$ other nodes. The codes presented for both these relaxations are obtained by modifying the `product-matrix' codes of~\cite{ourProductMatrix_short}.

We now take a brief digression to discuss a related notion, that of `repair-by-transfer', which shall be called upon frequently in the paper. Observe that when~\eqref{eq:mbr_eq_d} and~\eqref{eq:mbr_eq_r} are satisfied, the amount of download $\beta_D$ is equal to the amount of read $\beta_R$. As a result, whenever~\eqref{eq:mbr_eq_d} and~\eqref{eq:mbr_eq_r} are met, each of the $d$ nodes helping in the recovery must simply pass a part of the data that it stores, without performing any computations. This is termed \textit{repair-by-transfer}~\cite{ourAllertonJournal_short}. It follows that a repair-by-transfer code that satisfies~\eqref{eq:mbr_eq_d} for the amount of download automatically achieves~\eqref{eq:mbr_eq_r} for the read as well. Thus the problem considered in this paper can equivalently be stated as follows: \textit{for the MBR setting described above, under what conditions is it possible to design a code that can perform repair-by-transfer with a download satisfying~\eqref{eq:mbr_eq_d}?}

The rest of the paper is organized as follows. Section~\ref{sec:literature} describes related literature. Section~\ref{sec:non_ach} presents an information-theoretic proof showing the impossibility of achieving the previously derived lower bounds. Section~\ref{sec:relax} considers (practical) relaxations of this setting, and provides explicit codes operating under these relaxations. Section~\ref{sec:conclusions} presents conclusions.

\section{Related Literature}\label{sec:literature}
As described previously, explicit codes meeting~\eqref{eq:mbr_eq_d} and~\eqref{eq:mbr_eq_r} for recovery of the data of any node are presented in~\cite{ourAllerton_supershort,ourAllertonJournal_short} for the MBR setting when $d=n-1$. The notion of `repair-by-transfer' is also introduced therein. The repair-by-transfer codes of~\cite{ourAllerton_supershort,ourAllertonJournal_short} were subsequently extended to a more general but relaxed setting in~\cite{rouayheb2010fractional_short}. In~\cite{rouayheb2010fractional_short}, the condition of efficiently recovering the data of an individual node from \textit{any} $d$ nodes is relaxed to doing so from specific subsets $d$ nodes (termed `table-based' repair), with respect to both the amount of read and the amount of download. In contrast, the relaxations presented subsequently in this paper make such relaxations only for the amount of read, and the amount of download continues to achieve~\eqref{eq:mbr_eq_d} for every set of $d$ nodes.

In addition to the MBR setting discussed above, the regenerating codes model of~\cite{YunDimKanJournal_short} has another setting associated to it: the minimum storage regeneration (MSR) setting. Under the MSR setting, the storage is required to be at an absolute minimum, and for this value of storage, the amount of download is optimized. The problem of minimizing read in the MSR setting is studied in~\cite{ourMISERjournal_supershort,cadambe2011permutation_short,tamo2011mds_short,shum2012functional_short}. In particular, MSR codes performing repair-by-transfer with a minimum download for the systematic nodes are  constructed in~\cite{ourMISERjournal_supershort,cadambe2011permutation_short,tamo2011mds_short}. A somewhat different setting called `functional' repair is considered in~\cite{shum2012functional_short} for the application of repair of failed nodes. In this setting, the node replacing a failed node may recover data that is different from what was stored in the failed node, but which retains certain desired properties. MSR codes performing functional repair-by-transfer with minimum download for all nodes are constructed in~\cite{shum2012functional_short}.

\section{Impossibility of Repair-by-transfer in MBR when $d\neq n-1$}\label{sec:non_ach}
It was shown previously in~\cite{ourAllerton_supershort,ourAllertonJournal_short} that when $d=n-1$, both~\eqref{eq:mbr_eq_d} and~\eqref{eq:mbr_eq_r} can be achieved simultaneously. In this section, we present the converse to this result: we show that when $d\neq n-1$, there cannot exist any code under which the data stored in any node can be recovered from any $d$ other nodes while satisfying~\eqref{eq:mbr_eq_d} and~\eqref{eq:mbr_eq_r}. This result encompasses both linear and non-linear codes. The proof may be skipped without any loss in continuity.
\begin{thm}\label{thm:non_ach}
Under the MBR setting, when $d\neq n-1$, there cannot exist any code that performs repair-by-transfer of any node from any $d$ other nodes with a download satisfying~\eqref{eq:mbr_eq_d}.
\end{thm}
\begin{IEEEproof}
The proof proceeds via a contradiction. Let us suppose there exists such a code for some system parameters with $d\neq n-1$. The proof is divided into three parts. First, it is shown that there exist (at least) three nodes that store (at least) one bit of data in common. Next, it is shown that for recovery of the data of any one of these nodes, the other two nodes must pass this bit. Finally, we show that under this condition, such an attempt of recovery must necessarily fail.

\newcommand{\subD}{{}_\mathcal{D}\!}
For $i \in \{1,\ldots,n\}$, let $W_i$ be a random variable corresponding to the data stored in node $i$. For recovering the data of node $i$ from a set $\mathcal{D}$ of $d$ nodes, let $\subD S_j^i$ denote the random variable corresponding to the data passed by node $j\in\mathcal{D}$ to node $i$. Let $H(\cdot)$ denote Shannon entropy and $I(\cdot;\cdot)$ the mutual information. Let $\beta_D=\beta_R=\beta$. In the proof, we shall employ the following four properties, established in~\cite{ourAllertonJournal_short}, that any such code must satisfy. 

\cite[Property 1]{ourAllertonJournal_short} $H(W_i) = \alpha$

\cite[Property 2]{ourAllertonJournal_short} $I(W_i; W_j) = \beta$

\cite[Property 3]{ourAllertonJournal_short} $H(\subD S_j^i) = \beta$

\cite[Lemma 3]{ourAllertonJournal_short}~ $H(W_i|\subD S_j^i) \leq (d-1)\beta$\ ,\\\indent\hspace{2.2cm} $H(W_i|\subD S_j^i,\subD S_\ell^i) \leq (d-2)\beta$\ .

Consider recovery of the data stored in nodes $\{1,\ldots,d+1\}$ (one at a time), from node $n$ and $(d-1)$ other arbitrary nodes. In each case, node $n$ passes a {subset} of $\beta$ bits out of the $\alpha~(=d\beta)$ bits that it stores. We emphasize that due to the requirement of repair-by-transfer, the bits passed are simply \textit{subsets} of those it stores (and do not arise from any computations on the stored bits). Node $n$ thus passes a total of $(d+1)\beta$ bits. It follows from the pigeonhole principle that there exists at least one bit that occurs at least twice in this set of $(d+1)\beta$ bits. Moreover, \cite[Property 3]{ourAllertonJournal_short} implies that the $\beta$ bits passed by a node, for recovery of the data of any other node, must all be distinct. Thus there must exist at least two nodes out of $\{1,\ldots,d+1\}$ for which node $n$ passes the same bit. Let us assume that these two nodes are nodes $1$ and $2$, and let $b$ denote this common bit.

Since the data of any node $i$ must be completely recovered, the quantity $W_i$ is deterministic given the data passed by the $d$ nodes in the recovery process. It follows from\cite[Property 1]{ourAllertonJournal_short} that the entropy of the $d\beta~(=\alpha)$ bits passed by the $d$ nodes is $\alpha$. As a special case, it follows that $H(b)=1$. From the description above, one can also see that that $H(b|W_1)=0$ and $H(b|W_2)=0$. Moreover, since bit $b$ was originally stored in node $n$, $H(b|W_n)=0$. Thus, the bit $b$ is stored in nodes $1$, $2$ and $n$, and $H(b)=1$.

Now consider recovering the data of node $n$ from nodes $\{1,\ldots,d\}$. We shall now show that nodes $1$ and $2$ must both pass bit $b$. Abbreviating our earlier notation, we let $S_1$ and $S_2$ be random variables corresponding to data passed by nodes $1$ and $2$ respectively. Thus, by definition, we have $H(S_1|W_1)=H(S_2|W_2)=0$. From the properties discussed above, we get
\bea 
\!\!\!\!\!\!\!\!\!\!\!\!\!\!\!\!\!~~~~~~~~~2\beta\!\!\! &\!\!=\!\!&\! I(W_n;W_1) + I(W_n;W_2)\nonumber\\
&\!\!\geq\!\!&\! I(W_n; b, S_1) + I(W_n; b, S_2) \nonumber\\
&\!\!=\!\!& \!I(W_n;S_1) + H( b | S_1) - H(b| W_n, S_1)\nonumber\\
&& + I(W_n;S_2) + H( b | S_2) - H(b| W_n, S_2)\nonumber\\
&\!\!=\!\!&\! I(W_n;S_1) +\! H( b | S_1) +\! I(W_n;S_2) +\! H( b | S_2) \nonumber\\
&\!\!=\!\!& \!\!2H(W_n\!) \!-\! H(W_n|S_1\!) \!+\! H( b | S_1\!) \!-\! H(W_n|S_2\!) \!+\! H( b | S_2\!)\nonumber\\
&\!\!=\!\!& \!2d\beta\!-\! H(W_n|S_1) \!+\! H( b | S_1)  \!-\! H(W_n|S_2) \!+\! H( b | S_2) \nonumber\\
&\!\!\geq\!\!&\! 2d\beta - (d-1)\beta + H( b | S_1) - (d-1)\beta+ H( b | S_2) \nonumber\\
&\!\!=\!\!& \! 2\beta + H( b | S_1)+ H( b | S_2) ~.\nonumber
\eea

\noindent Thus, $ H(b|S_1) =H(b|S_2) = 0.$
It follows that
\bea 2\beta &=& H(W_n) - (d-2)\beta\nonumber\\
&\leq& H(W_n) - H(W_n|S_1, S_2) \nonumber\\
&=& I(W_n;S_1, S_2)\nonumber\\
&\leq& H(S_1, S_2)\nonumber\\
&\leq& H(S_1, S_2, b) \nonumber\\
&=& H(S_2) + H(b | S_2) + H(S_1 | S_2, b)\nonumber\\
%&=& \beta + 0 + H(S_1 | S_2, b)\nonumber\\
&\leq& \beta + 0 + H(S_1 | b)\nonumber\\
&=& \beta + H(b| S_1) + H(S_1) - H(b)\nonumber \\
&=& 2\beta - 1\label{eq:c}~.
\eea

\noindent 
Clearly,~\eqref{eq:c} yields a contradiction.
\end{IEEEproof}

\section{Explicit Codes for Two Relaxations} \label{sec:relax}
Repair-by-transfer under the regenerating codes setting described above amounts to (efficiently) recovering the contents of \textit{any} failed node from \textit{any} of the $d$ nodes. We saw in the previous section that the bounds of~\eqref{eq:mbr_eq_d} and~\eqref{eq:mbr_eq_r} \textit{cannot} be achieved simultaneously when $d\neq n-1$. Thus in this section, we consider two relaxations to this setup, which shall allow us to achieve these bounds. The two relaxations are obtained by slackening the two instances of the quantifier ``any'' for the amount of read. Note that under both relaxations, we shall continue to impose the requirements of the MBR setting, i.e., of recovering the entire data from any $k$ nodes, and satisfying~\eqref{eq:mbr_alpha_1} and~\eqref{eq:mbr_eq_d} on the amount of download for recovery of the data of any node from any $d$ nodes.

\subsection{Optimal recovery for systematic nodes}
A systematic code is defined as one under which some $k$ out of the $n$ nodes store data in a raw (uncoded) form. These $k$ nodes are called the systematic nodes, while the other $(n-k)$ nodes are termed parity nodes. For many applications such as degraded reads, efficient recovery of the data in a systematic node is of greater importance than that of a parity node. Keeping this in mind, we relax the setting described above to the following requirements: \begin{itemize}\item one should be able to recover the data stored in any node from any $d$ other nodes with a download equal to~\eqref{eq:mbr_eq_d} \item one should be able to recover the data stored in any systematic node, from any $d$ other nodes, with the read and download equal to~\eqref{eq:mbr_eq_r} and~\eqref{eq:mbr_eq_d} respectively.\end{itemize}
In other words, the requirement of repair-by-transfer is relaxed to hold only when recovering the data of a systematic node.

We now present an explicit code that achieves the conditions listed above. This code is a modification of the `product-matrix' MBR code of~\cite{ourProductMatrix_short}.~\footnote{While we discuss only the MBR case here, the ideas presented are also applicable to the product-matrix MSR codes of~\cite{ourProductMatrix_short}.} 
The code is linear, and operates over any finite field $\mathbb{F}_q$ of size $q~(\geq n)$. As in~\cite{ourProductMatrix_short}, we present constructions for the case when $\beta_D=1$ symbol over $\mathbb{F}_q$; codes for a general $\beta_D$ can be obtained via multiple concatenations of this code~(see \cite[Section I-C]{ourProductMatrix_short}). When $\beta_D=1$ symbol over $\mathbb{F}_q$,~\eqref{eq:mbr_alpha_1} reduces to having $\alpha=d$ symbols over $\mathbb{F}_q$.

We first present a brief overview of the construction of a product-matrix MBR code as in~\cite{ourProductMatrix_short}. Denote this code as $\mathcal{C}$. The product-matrix MBR code is designed to satisfy~\eqref{eq:mbr_eq_d}, i.e., when $\beta_D=1$  symbol over $\mathbb{F}_q$, it operates on a data of size 
\beq B=kd-{k \choose 2}\label{eq:B_total}\eeq
symbols over $\mathbb{F}_q$. Under the encoding mechanism of~\cite{ourProductMatrix_short}, this data is arranged as the entries of a $(d \times d)$ \textit{symmetric} matrix $M$ of the form
\[ M = \left[ \begin{array}{cc}S & R\\ R^T  &0\end{array} \right] \]
where $S$ is a $(k \times k)$ \textit{symmetric} matrix and $R$ is a $(k \times (d-k))$ matrix. $R^T$ denotes the transpose of $R$, and $0$ is a $((d-k)\times(d-k))$ zero matrix. Observe that the total number of independent entries in $S$ is $\frac{k(k+1)}{2}$ and that in $R$ is $k(d-k)$, and these two quantities add up to~\eqref{eq:B_total}.

Each node $i \in \{1,\ldots,n\}$ in the product-matrix MBR code is associated to a $d$-length vector $\boldsymbol{\psi}_i$. The vectors $\{\boldsymbol{\psi}_i\}_{i=1}^{n}$ are chosen to satisfy two conditions: (a) any $d$ of these $n$ vectors are linearly independent, and (b) when restricted to the first $k$ components, any $k$ of these $n$ vectors are linearly independent.

Every node $i\in \{1,\ldots,n\}$ stores the $\alpha~(=d)$ symbols \[\boldsymbol{\psi}_i^T M~.\] In this section, we shall assume that the code is systematic~\cite[Theorem 1]{ourProductMatrix_short},~\cite[Section IV-B]{ourProductMatrix_short} with nodes $\{1,\ldots,k\}$ being the systematic nodes.

It is shown in~\cite[Theorem 3]{ourProductMatrix_short}, by means of an explicit decoding algorithm, that the entire data can be recovered from the data of any $k$ of the $n$ nodes. This exploits the property of linear independence of the first $k$ components of $\{\boldsymbol{\psi}_i\}_{i=1}^{n}$.

Let us now look at recovering the data stored in an individual node $i\in\{1,\ldots,n\}$ from some $d$ nodes $\{j_1,\ldots,j_d\}$. Under the protocol proposed in~\cite{ourProductMatrix_short}, each of these $d$ nodes computes the inner product of the $d$ symbols stored in it with the $d$-length vector $\boldsymbol{\psi}_i$, and passes the result. Thus, the aggregate data obtained is $\{\boldsymbol{\psi}_{j_1}^TM \boldsymbol{\psi}_i,\ldots,\boldsymbol{\psi}_{j_d}^TM\boldsymbol{\psi}_i\}$. The linear independence of the $d$ vectors $\{\boldsymbol{\psi}_{j_1},\ldots,\boldsymbol{\psi}_{j_d}\}$ and the symmetry of matrix $M$ allows for recovery of the desired data $\boldsymbol{\psi}_i^T M$. Observe that the amount of download is equal to $d$ symbols over $\mathbb{F}_q$, and hence the code achieves~\eqref{eq:mbr_eq_d}.

We shall now modify the code $\mathcal{C}$ described above to obtain a new code $\mathcal{C}_1$ that, in addition, also minimizes the read during recovery of the data stored in any systematic node. Define a $(d \times d)$ matrix \beq\Psi_\sys = [\boldsymbol{\psi}_1~\boldsymbol{\psi}_2~\ldots~\boldsymbol{\psi}_d]~.\label{eq:psi_sys}\eeq Under $\mathcal{C}_1$, each node $i\!\in\!\{1,\ldots,n\}$ stores the $\alpha~(=d)$ symbols \[\boldsymbol{\psi}_i^TM\Psi_\sys\] (as opposed to storing $\boldsymbol{\psi}_i^TM$ under $\mathcal{C}$).
 
Let us now verify that code $\mathcal{C}_1$ meets all the requirements. First, observe that the $(d\times d)$ matrix $\Psi_\sys$ is invertible. Thus, the data stored in any node under $\mathcal{C}_1$ is equivalent~\cite[Appendix B]{ourProductMatrix_short} to that stored under $\mathcal{C}$. This results in the fulfilment of the conditions of recovery of the entire data from any $k$ nodes, and recovery of the data stored in any node from any $d$ nodes with a download equalling~\eqref{eq:mbr_eq_d}. 

Now consider recovering the data stored in any systematic node $i\in\{1,\ldots,k\}$ from any $d$ nodes $\{j_1,\ldots,j_d\}$. Under $\mathcal{C}_1$, every node $\ell \in \{j_1,\ldots,j_d\}$ simply reads and passes the $i^\supth$ symbol it stores, which from~\eqref{eq:psi_sys}, equals $\boldsymbol{\psi}_{\ell}^TM\boldsymbol{\psi}_i$. The data thus obtained is identical to that obtained under $\mathcal{C}$, thereby ensuring successful recovery. The amount of read and download is exactly $d$, thus meeting~\eqref{eq:mbr_eq_r} and~\eqref{eq:mbr_eq_d}.~\footnote{We note that this property, in fact, is applicable to the repair of any of the first $d~(\geq k)$ nodes.}

\subsection{Optimal recovery from $d$ specific nodes}
In certain applications, the flexibility of minimizing the read from \textit{any} set of $d$ nodes may be an overkill. This motivates the next relaxation, that mandates the following requirements: \begin{itemize} \item one should be able to recover the data stored in any node from any $d$ other nodes with a download equal to~\eqref{eq:mbr_eq_d} \item for any node, there must exist at least one set of $d$ other nodes such that recovery from these $d$ nodes entails a read and download equal to~\eqref{eq:mbr_eq_r} and~\eqref{eq:mbr_eq_d} respectively.\end{itemize} In other words, for recovery of the data stored in any node, the requirement of repair-by-transfer is relaxed to hold only for any one subset of $d$ nodes.

We now modify the product-matrix MBR code $\mathcal{C}$ described above to obtain a code $\mathcal{C}_2$ that satisfies these conditions. To simplify notation, define an operator $\oplus:\{1,\ldots,n\}\times\{1,\ldots,n\}\rightarrow\{1,\ldots,n\}$ that computes a sum that cycles in the set $\{1,\ldots,n\}$, i.e., for any $x,\ y \in \{1,\ldots,n\}$, $x\oplus y := 1+( (x-1 + y) \text{ mod } n)$. Let $\ominus$ be an analogous subtraction operator, with $x\ominus y := 1+((x-1-y) \text{ mod } n )$. Under $\mathcal{C}_2$, each node $i \in \{1,\ldots,n\}$ stores the $\alpha$ symbols 
\[\boldsymbol{\psi}_i^T M \left[ \boldsymbol{\psi}_{i\oplus 1} \ \boldsymbol{\psi}_{i\oplus 2} \ \cdots \  \boldsymbol{\psi}_{i\oplus d}\right]~.\] 

Let us now verify that code $\mathcal{C}_2$ meets all the requirements. Since any $d$ vectors from the set $\{\boldsymbol{\psi}_1,\ldots,\boldsymbol{\psi}_n\}$ are linearly independent, the matrix $\left[ \boldsymbol{\psi}_{i\oplus 1} \ \boldsymbol{\psi}_{i\oplus 2} \ \cdots \  \boldsymbol{\psi}_{i\oplus d}\right]$ is invertible for every $i$. Thus the data stored by a node under $\mathcal{C}_2$ is equivalent~\cite[Appendix B]{ourProductMatrix_short} to that stored under $\mathcal{C}$. This results in the fulfilment of the properties of recovery of the entire data from any $k$ nodes and recovery of data of any individual node from any $d$ nodes with a minimum download. 

Under $\mathcal{C}_2$, in order to recover the data stored in any node $i$ with a read and download equal to~\eqref{eq:mbr_eq_r} and~\eqref{eq:mbr_eq_d}, the $d$ nodes $(i\ominus d),\ldots,(i\ominus 1)$ are queried. Each node $\ell \in \{(i\ominus d),\ldots,(i\ominus 1)\}$ simply reads and transfers the symbol $ \boldsymbol{\psi}_\ell^T M \boldsymbol{\psi}_i$ that it has stored. The data thus obtained is identical to that obtained under $\mathcal{C}$, allowing for successful recovery of the desired data. This meets the bounds~\eqref{eq:mbr_eq_r} and~\eqref{eq:mbr_eq_d} on the read and download.  

\section{Conclusions}\label{sec:conclusions}
We consider the problem of constructing codes for distributed storage under which the data stored in any individual node can be efficiently recovered from the remaining nodes. In particular, we wish to achieve the previously derived~\cite{YunDimKanJournal_short} lower bounds on the amount of download and read. Achieving these bounds is equivalent to performing \textit{repair-by-transfer}~\cite{ourAllerton_supershort,ourAllertonJournal_short} while meeting the bound on the download. Explicit codes with these properties were constructed previously~\cite{ourAllerton_supershort,ourAllertonJournal_short} for $d=n-1$, and in this paper, we complete the picture by providing the converse to this result. In particular, we provide an information-theoretic impossibility result to show that the bound is not achievable when $d\neq n-1$ (even with non-linear codes). Obtaining tighter lower and upper bounds on the read under this setting are interesting directions for future research.

We also construct explicit codes for the two following (practical) relaxations, meeting the aforementioned bounds: (a) the read and download is simultaneously optimized for recovery of data of only systematic nodes, only the download is optimized for other nodes, and (b) for recovery of the data of any node, there is at least one set of $d$ nodes from which the read and download are simultaneously optimized, only the download is optimized for recovery from any other set of $d$ nodes. These codes are obtained by modifying the product-matrix codes of~\cite{ourProductMatrix_short}.

\bibliographystyle{IEEEtran}

\end{document}